\documentclass[useAMS,usenatbib]{mnras}
\usepackage{txfonts}
\usepackage{natbib}
\usepackage{graphicx}% Include Figure files
\usepackage{dcolumn}% Align Table columns on decimal point

\begin{document}

\title[On the GSD of turbulent dust growth]{On the grain-size distribution of turbulent dust growth}

\author[Mattsson]{Lars Mattsson$^{1}$\thanks{E-mail: lars.mattsson@su.se}\\
$^1$Nordita, KTH Royal Institute of Technology and Stockholm University, Roslagstullsbacken 23, SE-106 91 Stockholm, Sweden}

\pagerange{\pageref{firstpage}--\pageref{lastpage}} \pubyear{2020}

\maketitle

\label{firstpage}

\date{\today}

\begin{abstract} 
It has recently been shown that turbulence in the interstellar medium (ISM) can significantly accelerate the growth of dust grains by accretion of molecules, but the turbulent gas-density distribution also plays a crucial role in shaping the grain-size distribution. The growth velocity, i.e., the rate of change of the mean grain radius, is proportional to the local gas density if the growth species (molecules) are well-mixed in the gas. As a consequence, grain growth happens at vastly different rates in different locations, since the gas-density distribution of the ISM shows a considerable variance. Here, it is shown that grain-size distribution (GSD) rapidly becomes a reflection of the gas-density distribution, irrespective of the shape of the initial GSD. This result is obtained by modelling ISM turbulence as a Markov process, which in the special case of an Ornstein-Uhlenbeck process leads to a lognormal gas-density distribution, consistent with numerical simulations of isothermal compressible turbulence. This yields an approximately lognormal GSD; the sizes of dust grains in cold ISM clouds may thus not follow the commonly adopted  power-law GSD with index -3.5, but corroborates the use of a log-nomral GSD for large grains, suggested by several studies. It is also concluded that the very wide range of gas densities obtained in the high Mach-number turbulence of molecular clouds must allow formation of a tail of very large grains reaching radii of several microns.
\end{abstract}


\section{Introduction}
Grain growth in cold molecular clouds (MCs) from seed grains present at the formation of MCs is a scenario which has been generally accepted for for a long time \citep{Lindblad35,Baines65b,Baines65c}. This type of dust formation is an important dust-formation channel in many models of various redshifts and galaxy types \citep[see, e.g.,][]{Dwek98,Calura08,Mattsson11b,Valiante11,Asano13a,Ginolfi18} and depletion patterns in ISM gas are indeed consistent with dust depletion due to grain growth in MCs
\citep[see, e.g.,][]{Jenkins09,DeCia16,Mattsson19b}.

Models of grain growth in the ISM usually rely on an assumption that the exact gas-density field can be replaced with the mean density, i.e., a kind of ``mean-field approach'' \citep[as in the works of][]{Asano14,Hirashita19,Aoyama20}. Unfortunately, this ``erases'' smaller scale variations and other effects of dynamics. However, if the density variations are sufficiently small, this is a reasonable approach. But in case of strong compressible turbulence the gas density can vary by orders of magnitude and a significant fraction of the molecular gas in an MC display densities well above the critical density required for efficient grain growth \citep[see][for more details about this critical density]{Asano13a}.

In a homogeneous (constant density) environment, grain-growth by accretion is mainly limited by the abundance of the growth-species molecules, which in turn is limited by the overall metallicity in the ISM. Thus, for a given metallicity, the gas density is decisive for the rate of grain growth. This is indicating that modelling grain growth in the ISM in terms of a locally constant mean density may be an incorrect approach. The cold molecular phase of the ISM is highly inhomogeneous and display strong gas-density variations on sub-parsec scales. Such gas-density variations mean that some regions have number densities of growth species which are high enough to reach very fast grain growth. Moreover, since gas and dust tend to be coupled (at least on average), a majority of dust grains may actually reside in those regions. In a recent paper, \citet[][henceforth M20]{Mattsson20} showed that the overall rate of accretion can be increased by as much as two orders of magnitude when turbulent density variations are taken into account. Numerical simulations of interstellar turbulence 
\citep[e.g.,][]{Klessen00,Price11,Konstandin12,Federrath13,Nolan15} have demonstrated a direct relationship between the gas-density variance and the root-mean.square Mach number $\mathcal{M}_{\rm rms}$, which gives that a relationship between $\mathcal{M}_{\rm rms}$ and the effective grain-growth velocity must exist. Hence, a significantly accelerated growth rate is expected in supersonic turbulence (see M20), which may resolve the apparent timescale crisis for dust formation at high redshifts \citep{Mattsson11a,Rowlands14a,Watson15}.

In a recent set of simulations by Li \& Mattsson (2020, submitted) it was seen that turbulence accelerated growth of dust grains can have also another effect: grains at different locations may experience different overall growth velocities. Consequently, the grain-size distribution may undergo more complex evolution than predicted by simple models \citep[see, e.g.,][]{Hirashita11,Mattsson16}. The aim of the present paper is to show how density variations due to strong ISM turbulence can be a key factor in shaping the grain-size distribution.

\section{Theory and method}
\label{theory}
\subsection{Dust growth by accretion}
\label{dustgrowth}
In case dust growth by accretion of specific molecules takes place in a homogeneous medium, i.e., a gas of constant density, the mathematical description of this process is rather simple. The GSD $f$ used here, can be seen as a probability density function if divided by total number density of grains $n_{\rm d}$, so that $f = n_{\rm d}^{-1}(dn/da)$, and must satisfy a ``continuity equation'' of the form
\begin{equation}
\label{eoc}
{\partial f\over \partial t}+ \xi(t)\, {\partial f\over \partial a} = 0,
\end{equation}
where $a$ is grain radius and $t$ is time. The thermal growth velocity $\xi$ is the rate by which $a$ increases  due to thermal collisions (and chemical reactions) with the considered molecular growth species. It is easy to show that $\xi$ is independent of $a$ and given by  \citep[see, e.g.][]{Hirashita11,Mattsson14a},
\begin{equation}
\label{grovel}
\xi(t) = S\,\bar{u}_{\rm t}\,X_i(t)\,{\rho(t)\over  \rho_{\rm gr}},
\end{equation}
where $X_i$ is the mass fraction of the relevant growth-species molecules $i$ in the gas, $S$ is the sticking probability for a molecule hitting the grain and $\bar{u}_{\rm t}$ is the thermal mean speed of the molecules (which is assumed to be constant).

A linear transformation of the form
\begin{equation}
A(a,t) = a - \langle a\rangle + a_0= a-\int_0^t\xi(t')\,dt',
\end{equation}
where $a_0$ is the initial mean radius, gives the formal solution to equation (\ref{eoc})
\begin{equation}
\label{formal}
f(a,t) = f_0\left[A(a,t)\right] = f_0\left[a-\int_0^t\xi(t')\,dt'\right],
\end{equation}
where $f_0$ is the GSD at $t=0$. The choice of $f_0$ is, from a mathematical point of view, completely arbitrary, but any reasonable GSD must be skewed towards the small-grain end \citep{Mattsson16}.

\subsubsection{Evolution of the GSD}
\label{evolution}
The formal solution can mathematically be classified as a translational-invariant function, a mapping of the form $f: \mathbb{R}^2\to\mathbb{R}$, where in this particular case the plane formed by time $t$ and grain radius $a$ can be projected onto a line (in physics often referred to as a ``travelling-wave solution''). In case the distribution $f$ is not changing its shape one may say that it has a {\it purely translational} evolution. A linear transformation of the form $\eta = a + b$ will in such a case merely shift $f$ to the left if $b>0$ and to the right if $b<0$. If only accretion is considered, the translation must be to the right ($b<0$).  

\begin{figure*}
  \resizebox{\hsize}{!}{
   \includegraphics{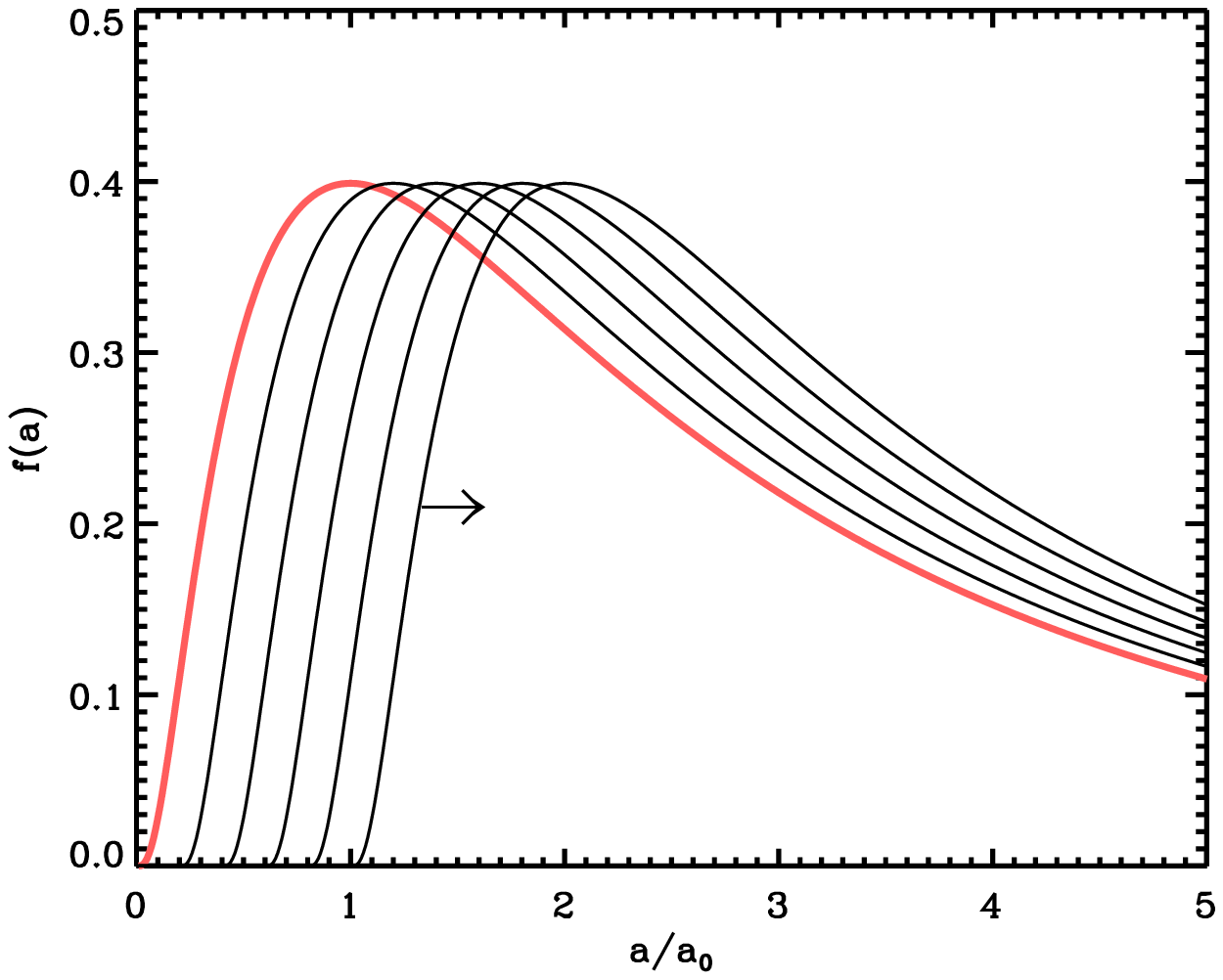}
   \includegraphics{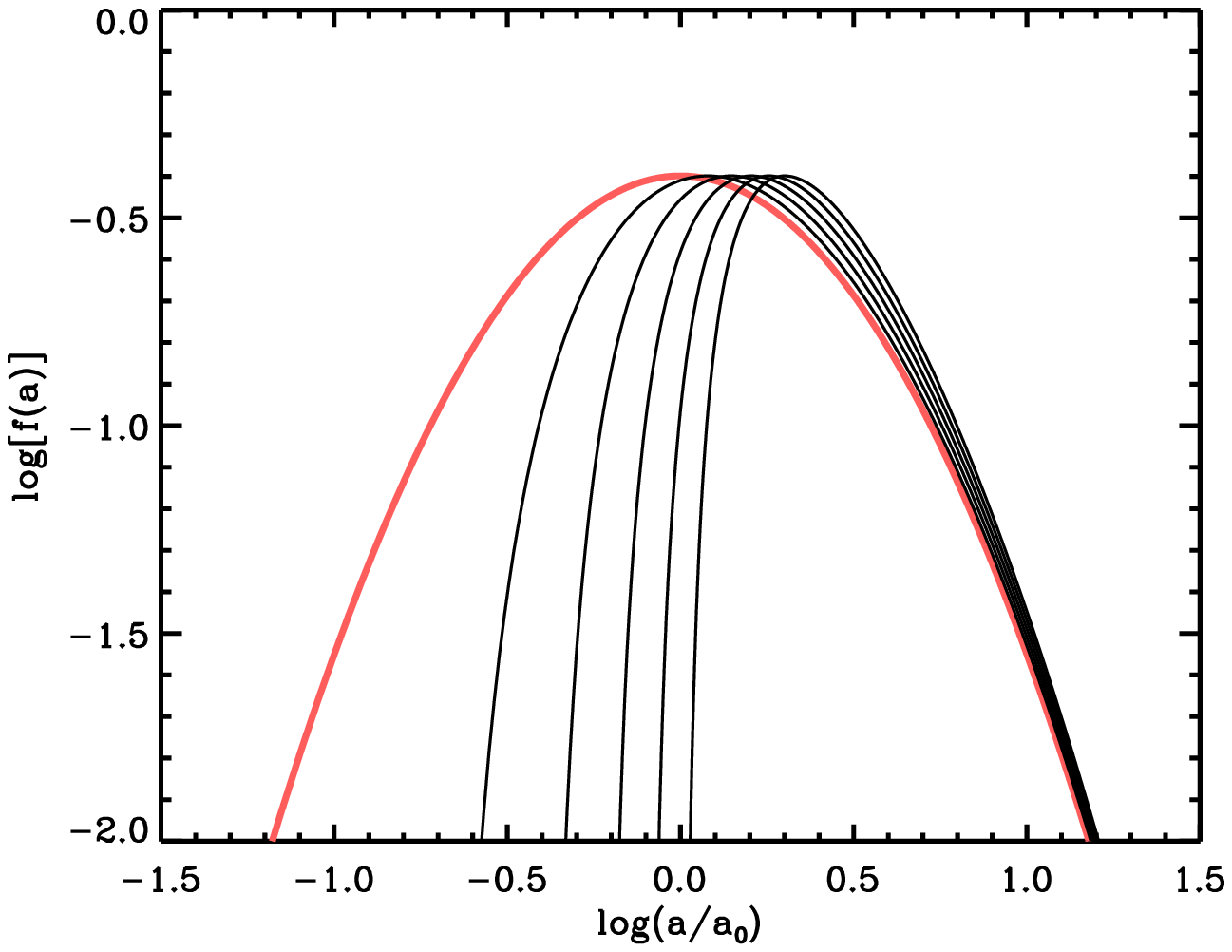}}
  \caption{\label{fsgsd} Evolution of an initially lognormal GSD assuming the growth velocity $\xi$ is independent of grain size. The GSD is simply moving to the right (left panel ), which appear as ``steepening'' of the GSD when plotted on a log scale (right panel). The reason for this is that the relative growth of small grains is faster than the growth of larger grains.}
  \end{figure*}

The left panel in Fig. \ref{fsgsd} shows how an initially lognormal GSD, $f_0(a) \propto \exp\{-{1\over 2}[\ln(a)-\langle \ln(a)\rangle]^2/\sigma_a^2\}$, is shifted to the right as the grains grow, while the right panel shows the apparent steepening of the GSD on a log-scale. In the literature it is sometimes said that the GSD becomes steeper and skewed towards the small-grain end \citep[see, e.g.,][]{Hirashita11}, which is only true in a relative sense. The relative increase of the radius $\Delta a/a$ of a small grain is much larger than $\Delta a/a$ for an initially large grain, which becomes apparent when plotted on a log-scale.

\begin{figure}
  \resizebox{\hsize}{!}{
   \includegraphics{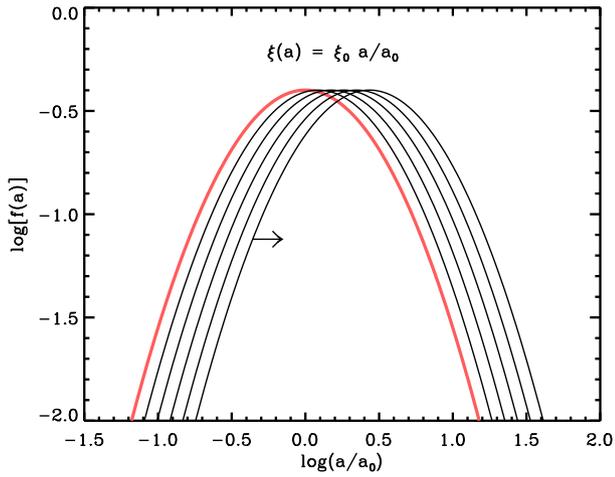}}
  \caption{\label{fsgsd_logtrans} {Evolution of an initially lognormal GSD assuming $\xi \propto a$. In this case the GSD is simply moving to the right when plotted on a log scale, which means its variance is actually increasing.}}
  \end{figure}

\begin{figure}
  \resizebox{\hsize}{!}{
   \includegraphics{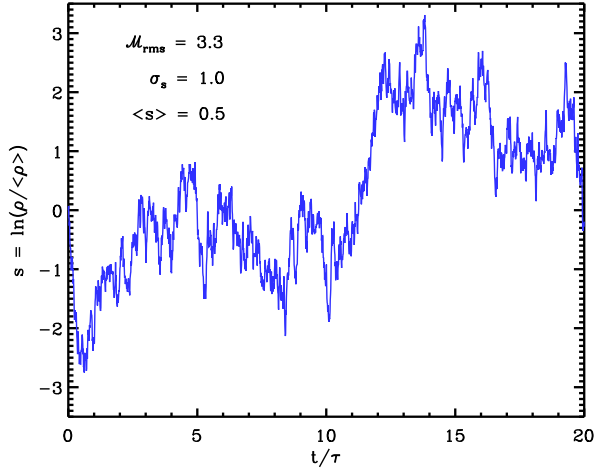}}
  \caption{\label{timeseries} Example of a time series of the evolution of the density $\rho$ of a fluid volume element as modelled by an Orstein-Uhlenbeck process. The timescale (unit) $\tau$ is the relaxation time of the process and the density $\rho$ is renormalised such that $\langle \rho\rangle$ is the mean also when taken over the time series.}
  \end{figure}

However, if $\xi$ is also a function of $a$, the evolution of the GSD may not be translational on a linear scale any more. In particular, with $\xi(a,t) = \xi_0(t)\,a/a_0$, the evolution of the GSD can be described by the equation 
\begin{equation}
\label{eoc_log}
{\partial \varphi\over \partial t}+ \xi_0(t)\, {\partial \varphi\over \partial \ln a} = 0,
\end{equation}
where $\varphi(a,t) = a/a_0\,f(a,t)$. This equation is solved by
\begin{equation}
\label{formal_log}
f(a,t) = f_0\left[\ln(a) - {1\over a_0}\int_0^t\xi_0(t')\,dt'\right],
\end{equation}
which is a solution that will appear translational-invariant on a log-scale. An example of this is shown i Fig. \ref{fsgsd_logtrans}, where a lognormal distribution is evolved assuming $\xi\propto a$. In such a case, the GSD actually evolves towards being strongly dominated by large grains. This situation could effectively occur if $\xi$ displays significant variance in space and time, in which case there is a certain probability that a grain will grow by a certain amount on an arbitrary time interval $t+\Delta t$. Grains which happen to be in regions of high $\xi$ (presumably with a high gas density $\rho$) will rapidly become large and may continue to grow faster than other grains which do not become large and continue to grow slowly because they reside in regions of low $\xi$ ($\rho$, or $\xi$, in a fluid element can remain below or above the mean for an extended time, as seen in the time-series example in Fig. \ref{timeseries}). Thus, the effective (integrated) $\xi$ may be increasing with $a$.

\subsection{Statistics of a turbulent gas}
\label{gasPDF}
Interstellar gas is turbulent and highly compressible. Many numerical simulations as well as observational studies suggest root-mean-square Mach numbers $\mathcal{M}_{\rm rms} \gtrsim 10$ in MCs \citep[e.g.,][]{Brunt10,Price11,Molina12,Nolan15}, which means the turbulence in the cold ISM is strong, highly compressible and displays a wide range of gas densities.

Numerical simulations of isothermal hydrodynamic turbulence with rotation-dominated forcing is known to produce roughly lognormal gas-density statistics \citep[see, e.g.,][and references therein]{Federrath10,Mattsson19a}.  Magneto-hydrodynamic simulations also yield roughly lognormal statistics \citep[see., e.g.,][and references therein]{Molina12}, but with suppressed density variance for very high-$\mathcal{M}_{\rm rms}$ turbulence \citep{Ostriker01,Price11}. However, it is the low-density tail that tend to be suppressed \citep{Molina12}, which means that the effect on processes mainly taking place in high-density regions (e.g., dust growth by accretion of molecules) is small. 

The lognormal distribution is of the form
\begin{equation}
\mathcal{P}(s) = {1\over \sqrt{2\pi}\,\sigma_{\rm s}}\exp\left[ -{(s - \mu)^2\over 2\,\sigma_{\rm s}^2}\right], \quad s = \ln\left({\rho\over\langle\rho\rangle}\right), 
\end{equation}
where $\mu$ is related to the variance/standard deviation such that mass conservation is obtained \citep{Vazquez94,Konstandin12}.

The variance can be estimated from its relation to the root-mean-square Mach number $\mathcal{M}_{\rm rms}$, usually considered to be of the form
\begin{equation}
\label{sigmamach}
\sigma_{\rm s}^2 = \ln(1+b^2\mathcal{M}_{\rm rms}^2),
\end{equation}
which has been confirmed by several numerical experiments \citep[e.g.,][]{Passot98,Federrath10}. A typical value for the case of for purely solenoidal forcing is $b=1/3$. For mixed forcing, a value $b\approx 0.5$ is often quoted \citep{Federrath13}. In the simulations described below, it will be assumed that $b = 0.4$ and the quoted $\mathcal{M}_{\rm rms}$ values are based on this assumption.

\subsection{Turbulence as a Markov process}
\label{markov}
Numerical simulations of particle-laden strongly turbulent gas where the particles are interacting with each other or the gas are computationally expensive and it can be difficult to identify the mechanisms behind emergent phenomena, such as the broadening of the GSD seen by Li \& Mattsson (2020, submitted). It may thus be useful to try a more analytical approach to test the hypothesis outlined in previous sections. Here, the density variations due to turbulence will therefore be modelled as a stochastic process and not by solving the equations of fluid dynamics.

It is well-established that the velocity field of a turbulent fluid can be described as a Markov process \citep[see, e.g.,][]{Novikov89,Pedrizzetti94} and the gas-density PDF in highly compressible turbulence has recently been modelled in a similar way \citep{Mocz19,Scannapieco18}. In the latter case, the approach is to describe turbulence in terms of the statistical time-evolution of fluid-element densities. The gas-density PDF can be regarded as made up by volume or mass elements and turbulence as a temporally homogeneous Markov process in which the future state space only depends on its current values. If the logarithmic density parameter $s$ (see section \ref{gasPDF}) is an Ornstein-Uhlenbeck (OU) process \citep{Uhlenbeck30}, a special type of Markov process to be described below, the resultant gas-density PDF $\mathcal{P}(\rho)$ is a lognormal distribution.

A Markov process $R$ of OU type, essentially a mean-reverting random walk, will obey a \citet{Langevin08} equation of the form
\begin{equation}
\label{langevin}
R(t+dt) - R(t) = {\langle R\rangle - R(t)\over \tau}\,dt + \mathfrak{D}^{1/2}\,\mathcal{N}(0,1)\,(dt)^{1/2},
\end{equation}
where $\tau$ is the relaxation time, $\mathfrak{D} = 2\sigma_s^2/\tau$ is a diffusion constant and $\mathcal{N}(\mu, \sigma^2)$ denotes a temporally uncorrelated Gaussian random variable with mean $\mu$ and standard deviation $\sigma$. The above equation can be equivalently written as a stochastic differential equation with a noise term,
\begin{equation}
\label{langevin2}
{dR\over dt} = {\langle R\rangle - R(t)\over \tau} + \mathfrak{D}^{1/2}\,\gamma(t),
\end{equation}
where $\gamma(t)$ is Gaussian white noise with $\mu =0$ and $\sigma = 1/dt$ in the present case. The equivalence of the two equations (\ref{langevin}) and (\ref{langevin2}) is a simple consequence of the property $\mathcal{N}(\alpha+\beta\,\mu, \beta^2\,\sigma^2) = \alpha + \beta\,\mathcal{N}(\mu, \sigma^2)$ of a Gaussian random variable.

The Markov process described above has a steady-state solution for the PDF of $R$. The PDF $\mathcal{P}(R,t)$ obeys the Fokker-Planck equation,
\begin{equation}
\label{fokker}
{\partial \mathcal{P}(R,t)\over \partial t} = {1\over \tau}{\partial \over\partial R}[R\,\mathcal{P}(R,t)] + {\mathfrak{D}\over 2}{\partial^2 \mathcal{P}(R,t)\over \partial R^2},
\end{equation}
which for a steady state ($\partial P/\partial t = 0$) reduces to a simple ordinary differential equation,
\begin{equation}
\label{fokker2}
{d^2P(R)\over dR^2} +  {d\over dR}\left[{(R-\langle R\rangle)\,P(R)\over  \sigma_s^2}\right] = 0,
\end{equation}
to which the only nontrivial solution is a Gaussian distribution. The time it takes for an OU process to reach this steady-state distribution is about twice\footnote{The conditional mean $\langle R\rangle$ of an OU process evolves towards a steady-state value via an exponential decay, where the relaxation time $\tau$ is the e-folding timescale \citep{Karatzas91}. After 2.3 e-folding times, $\langle R\rangle$ has moved 90\% of the distance between its initial and final values, regardless of what those values are, which is here considered to be sufficiently close to the (final) steady state to say the system is in the steady-state phase.} the relaxation time $\tau$.

\subsection{Numerical implementation of an Ornstein-Uhlenbeck process}
The first term on the right-hand side of the Langevin equation (\ref{langevin}) is a deterministic term (in $dt$), while the second term is a stochastic term. The infinitesimal step is a Gaussian random variable and the derivative is white noise, as described above. Numerical treatment of an OU process is commonly based on the Euler-Maruyama method, which
involves discretising time and adding infinitesimal steps to the process at every time step $\delta t$ \citep[see][]{Kloeden11}. That is, for the Langevin equation above, the scheme is
\begin{equation}
R_{n+1} =R_n+\delta R = R_n+ {\langle R\rangle - R_n\over \tau}\,\delta t+\mathfrak{D}^{1/2}\mathcal{N}(0,1)\,(\delta t)^{1/2},
\end{equation}
where all parameters are as previously defined. $\mathcal{N}(0,1)$ is a random Gaussian variable with mean $\mu = 0$ and variance $\sigma^2 = 1$, which is independent at each time step. The factor $(\delta t)^{1/2}$ arises from the fact that the incremental step for white noise must have $\sigma = (\delta t)^{1/2}$. Consequently, the error of the Euler-Maruyama method is of order $(\delta t)^{1/2}$. The scheme outlined above is a {first order} scheme and it should be noted that higher-order schemes can also be used \citep{Gillespie96}.

\subsection{Modelling a dust-gas system}
In order to simulate a dust-gas system a few of simplifying assumptions have to be made. Dust grains are inertial particles, which means they can decouple from the carrier fluid (the interstellar gas) if the frictional drag force acting on the grains is low enough relative to the inertial forces of the grains. As explained in M20, dust-gas coupling in turbulence is a both physically and mathematically complex problem, which requires advanced numerical simulation. To make the problem (grain growth by accretion) analytically tractable it is more or less necessary to assume that dust and gas are position coupled on the spatial scale of interest. Thus, a grain residing in a given fluid element at $t=0$ will reside in the same fluid element at any later time. Another assumption to be made is that the initial mass fraction of the growth-species molecules in the gas is a ``universal'' constant, i.e., the same for every fluid element. Effects of magnetic fields are here considered unimportant, but it should be noted in passing that the Lorentz force acting upon the grains may lead to {\it more efficient} coupling between gas and dust.
Moreover one may also assume that the thermal mean speed of the gas particles/molecules is everywhere the same; the modelled system is assumed to be strictly isothermal. This assumption means that the growth velocity $\xi$ will only vary due to variations in the gas density $\rho$.

\begin{table*}
\caption{Parameters of the OU simulations used in the present paper. All parameters are dimensionless except the grain radii, $a_0$, $a_{\rm min}$ and $a_{\rm max}$, where the latter two are the initial minimum and maximum radii of the grains.}
%\centering
\setlength{\tabcolsep}{5pt}
\begin{tabular}{ccccccccccccl}
Run & $N_{\rm trials}$ & $\delta t/\tau$ &$\cal{M}_{\rm rms}$ & $\sigma_s$ & $\langle s \rangle$ & $a_0$ & $a_{\rm min}$&$a_{\rm max}$ & $\beta$ & $\rho_{{\rm d}\,i}(0)/\rho_i(0)$ & Remark\\
 & & & & & &  ($\mu$m) &  ($\mu$m) &  ($\mu$m) & & & \\
\hline
1 & $10^5$ & 0.01 & 0.6 & 0.25 & 0.03125 & 0.01 & 0.01 & 0.01 & - & - & No depletion\\
2 & $10^5$ & 0.01 & 1.3 & 0.5 & 0.125 & 0.01 & 0.01 & 0.01 & - & - & No depletion\\
3 & $10^5$ & 0.01 & 3.3 & 1.0 & 0.5 & 0.01 & 0.01 & 0.01 & - & - & No depletion\\
4 & $10^5$ & 0.01 & 7.3 & 1.5 & 1.125 & 0.01 & 0.01 & 0.01 & - & - & No depletion\\
5 & $10^5$ & 0.01 & 18 & 2.0 & 2.0 & 0.01 & 0.01 & 0.01 & - & - & No depletion\\[3mm]

6 & $10^5$ & 0.01 & 0.6 & 0.25 & 0.03125 & 0.01 & $5.4\, 10^{-3}$ & 1.0 & 3.5 & - & No depletion\\
7 & $10^5$ & 0.01 & 1.3 & 0.5 & 0.125 & 0.01 & $5.4\, 10^{-3}$ & 1.0 & 3.5 & - & No depletion\\
8 & $10^5$ & 0.01 & 3.3 & 1.0 & 0.5 & 0.01 & $5.4\, 10^{-3}$ & 1.0 & 3.5 & - & No depletion\\
9 & $10^5$ & 0.01 & 7.3 & 1.5 & 1.125 & 0.01 & $5.4\, 10^{-3}$ & 1.0 & 3.5 & - & No depletion\\
10 & $10^5$ & 0.01 & 18 & 2.0 & 2.0 & 0.01 & $5.4\, 10^{-3}$ & 1.0 & 3.5 & - & No depletion\\[3mm]

11 & $10^6$ & 0.005 & 3.3 & 1.0 & 0.5 & 0.01 & 0.01 & 0.01 & - & 0.1 & \\
12 & $10^6$ & 0.005 & 3.3 & 1.0 & 0.5 & 0.01 & $5.4\, 10^{-3}$ & 1.0 & 3.5 & 0.1 & \\
\hline
%\multicolumn{13}{p{0.6\textwidth}}{}
\end{tabular}
\label{parameters}
\end{table*}

In the present study, following the discussion above, it is the growth velocity $\xi$ that is the stochastic process of interest. As mentioned above in section \ref{dustgrowth}, one may assume $\xi = \xi_0\, \rho/\langle \rho \rangle$, with $\xi_0$ a constant, for as long as depletion of the growth species is not significant. Thus, the stochastic variation of $s = \ln(\rho/\langle \rho \rangle)$ and $\ln(\xi/\xi_0)$ is actually governed by the same OU process, albeit with the addition of an ``amplitude correction'' due to depletion of the growth species. Discretising time in equation (\ref{eoc}) and introducing $s$ as a discrete OU process in that equation and tracking a large number $N_{\rm e}$ ``fluid elements'', one obtains a system of the form
\begin{eqnarray}
\label{system}
\nonumber
f_n(a_\ell) &=& {\rm HIST}[\ell, \Delta \log a] \left(a_0 + \sum_{k=0}^n \xi_{k} \right),\\
\xi_n &=& \xi_0\, \exp(s_n),\\
\nonumber
s_{n+1} &=& s_n+ {\langle s\rangle - s_n\over \tau}\,dt+\sigma_s\left({2\over \tau}\right)^{1/2}\mathcal{N}(0,1)\,(dt)^{1/2},
\end{eqnarray}
where $f_n$ is the discretised/binned GSD after $n$ time steps and $\xi_0$ is the growth velocity at the initial time step $(t = 0)$. The GSD evolution is obtained by binning the elements at all (or selected) time steps by applying the histogram function HIST$[\ell, \Delta \log a](\dots)$, where $\ell$ is the number of bins and $\Delta \log a$ is the logarithmic bin size. $N_{\rm e}$ must be large to ensure proper statistics and generally $N_{\rm e} \gg N$ and $N\gg\ell$.

To include the effect of depletion one can add a depletion factor $F^\star$ to the system of equations above. More exactly, the growth velocity-equation should then be replaced with
\begin{equation}
\label{depletion}
\xi_n = \xi_0\, F_n^\star\,\exp(s_n), \quad  F_n^\star = 1 - \exp(s_{\rm d} -s_i),
\end{equation}
where $\rho_{\rm d}$ is the local dust mass density, $s_{\rm d} = \ln(\rho_{\rm d}/\langle \rho\rangle)$ and $s_i = s  \ln[X_i(0)]$, {following the model of depletion in \citet{Mattsson16}}.

\section{Results and discussion}

\subsection{Numerical simulation of GSD evolution}
\label{numsim}
By numerical solution of the system of equations (\ref{system}) using the Euler-Maruyama method, the evolution of large number $N_{\rm e}$ of ``fluid volume elements'' can be followed and the evolution of the GSD can be reconstructed by binning the sizes of the grains associated with each element. In the simulations presented here $N_{\rm e} = 10^5$ and the initial sizes of the grains associated with each fluid element is either the same for all elements (``delta-distributed'' case) or randomly drawn from a power-law distribution with a slope $-3.5$ \citep[``MRN'' case, see][]{Mathis77}. Each time series is computed with a constant time step $\delta t/\tau = 0.01$ and the length of the time series is 20 relaxation times, i.e., $t_{\rm max} = 20\,\tau$. That is, the density evolution of each one of the $10^5$ fluid elements is modelled by a random walk with 2000 steps.

To explore what role the strength of the turbulence plays, simulations has been performed with different $\sigma_s$ (see Table \ref{parameters}) corresponding to a wide range of $\mathcal{M}_{\rm rms}$ values. For convenience, the unit time is equal to the relaxation time of the OU process, i.e., all simulations are made with $\tau = 1$ (arbitrary system of units). The lognormal gas-density PDF is assumed to be mass weighted, which requires that $\langle s \rangle = {1\over 2}\,\sigma_s^2$ to ensure mass conservation \citep{Vazquez94,Scannapieco18}. The parameters governing the OU process are dimensionless and the initial/average growth velocity $\xi_0$ is the only parameter that requires physical scaling in order to obtain grain sizes in physical units. In units of $\tau$, the adopted value in all simulations presented here is $\xi_0\,\tau = 10^{-3}\,\mu$m. Following equation (\ref{grovel}), an estimate of $\xi_0$ can be made from 
\begin{equation}
    \xi_0 = S\,u_{\rm therm}\, X_i(0)\,{\langle\rho\rangle\over \rho_{\rm gr}},
\end{equation}
where $S$ is the sticking probability, $u_{\rm therm}$ and $X_i(0)$ are the mean thermal velocity of the relevant molecules and their fraction of the total gas mass, respectively, and $\rho_{\rm gr}$ is the bulk material density of the grains. Assuming a mean density of $\langle\rho\rangle \sim 10^{-21}$~g~cm$^{-3}$, $u_{\rm therm}\sim 0.1$~km~s$^{-1}$, $X_i(0) \sim 0.01$, $\rho_{\rm gr}\approx 3$~g~cm$^{-3}$ and $S\approx 0.3$, the initial growth velocity is of the order $\xi_0 \sim 10^{-25}$~cm~s$^{-1}$. The assumption $\xi_0\,\tau = 10^{-3}\,\mu$m then implies $\tau \sim 1$~Myr in physical units.

  \begin{figure*}
  \resizebox{\hsize}{!}{
   \includegraphics{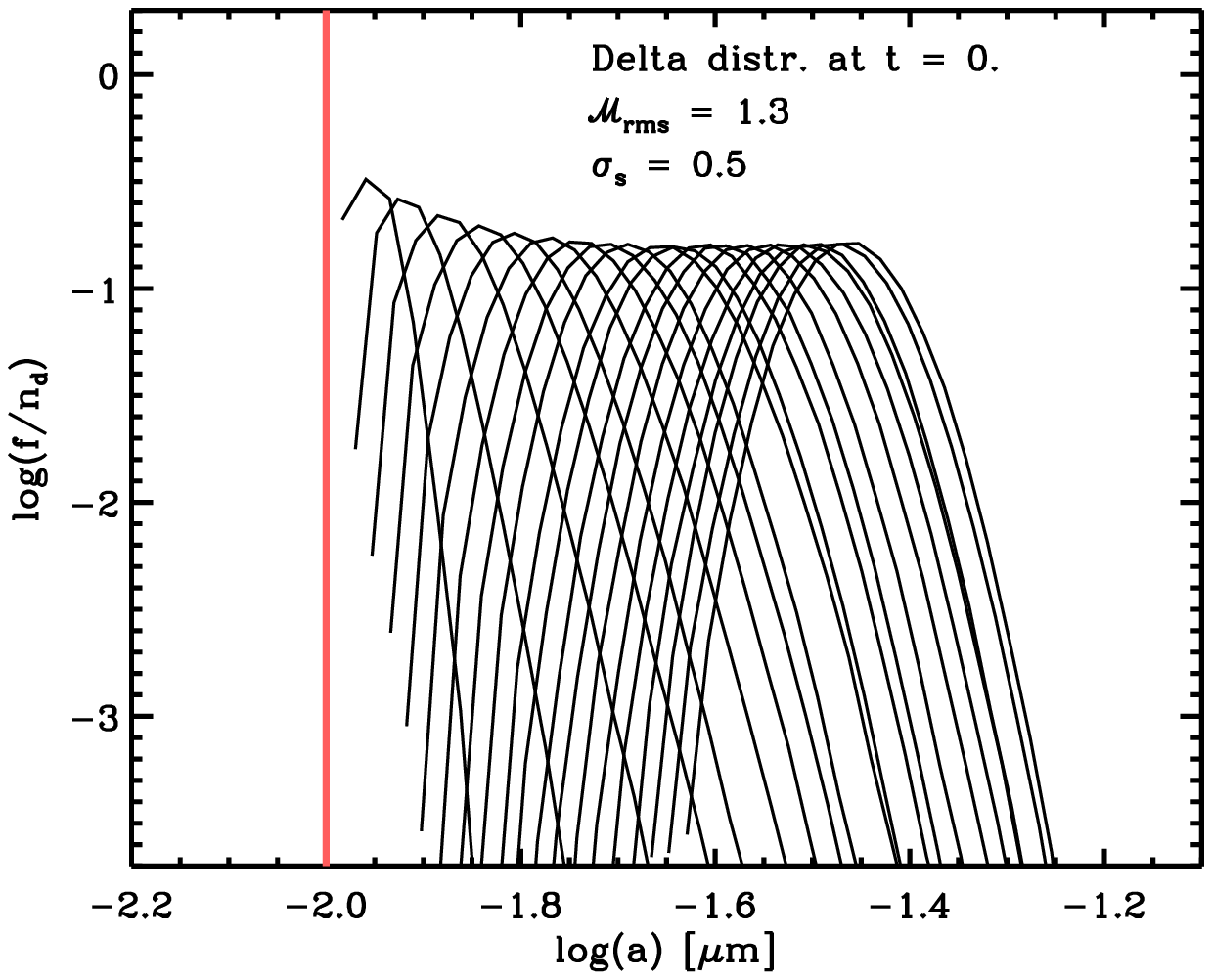}
  \includegraphics{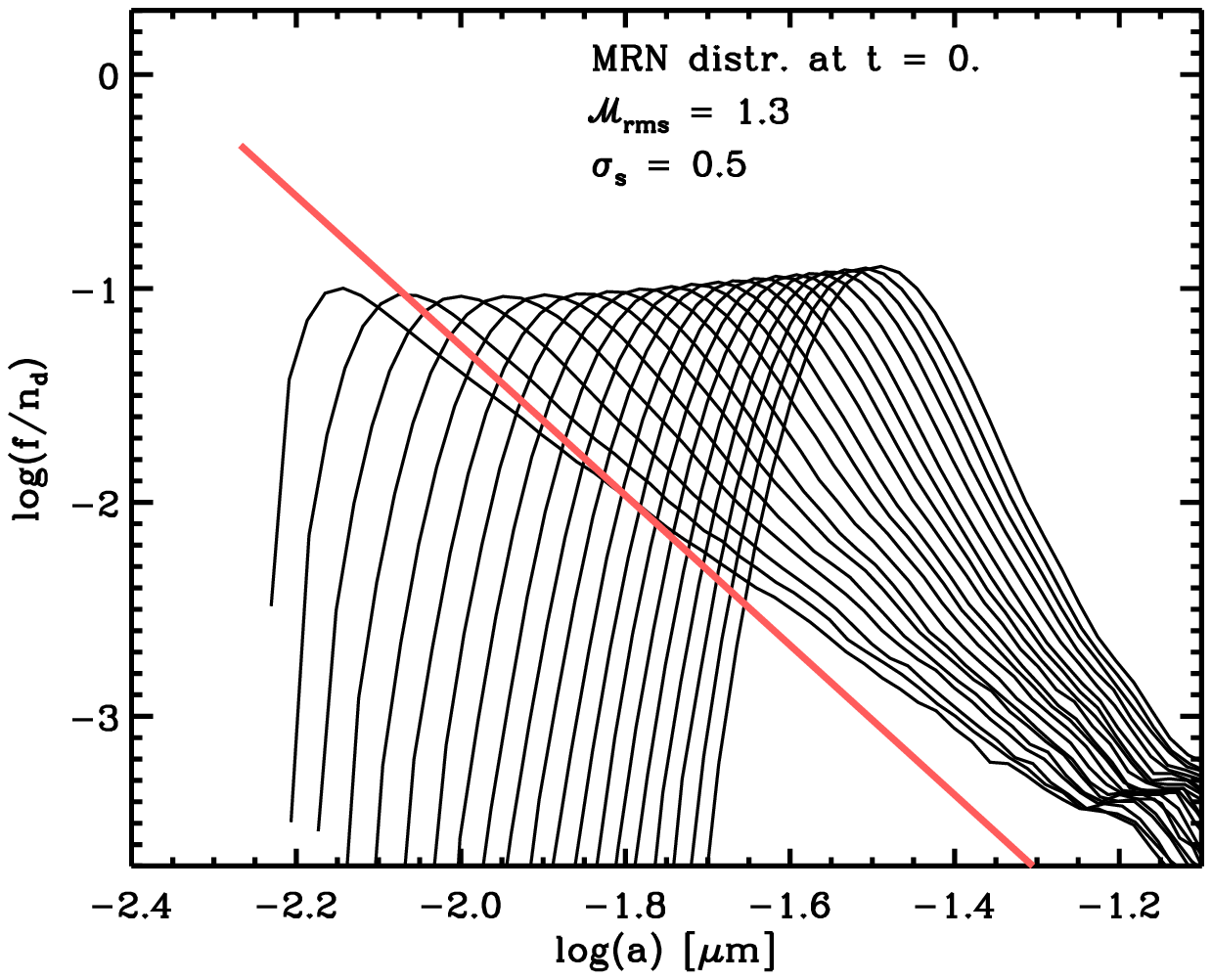}}
  \resizebox{\hsize}{!}{
   \includegraphics{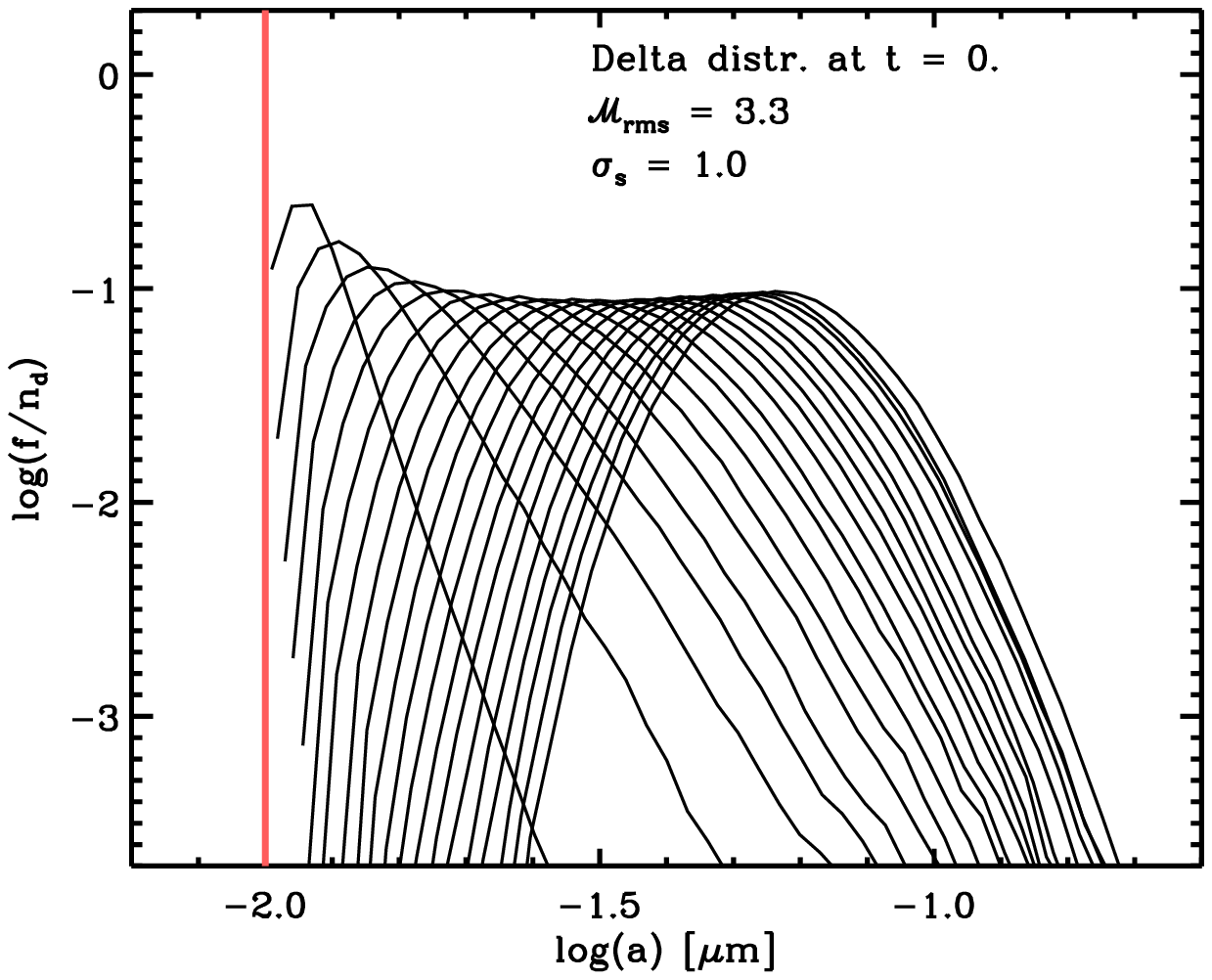}
  \includegraphics{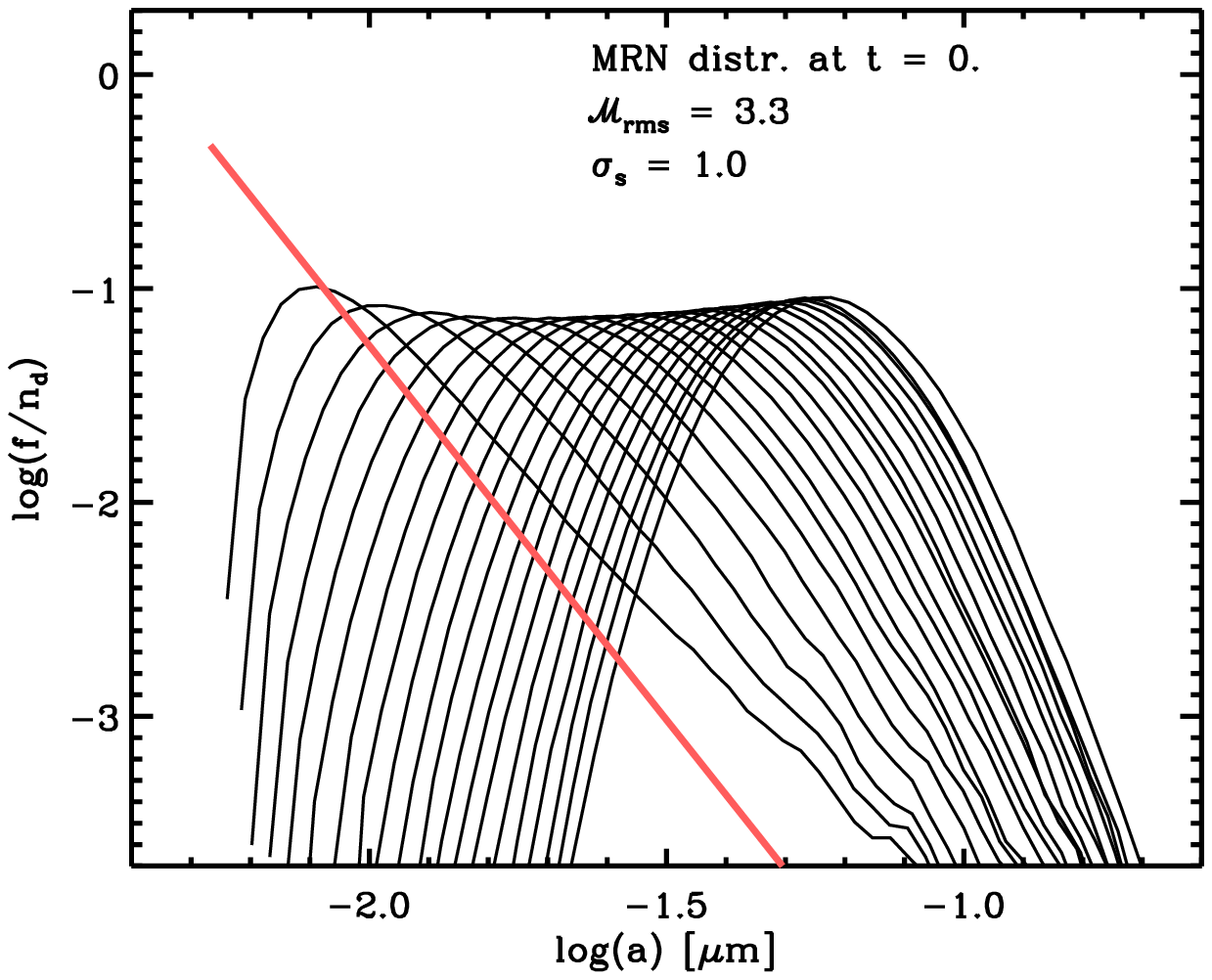}}
  \resizebox{\hsize}{!}{
   \includegraphics{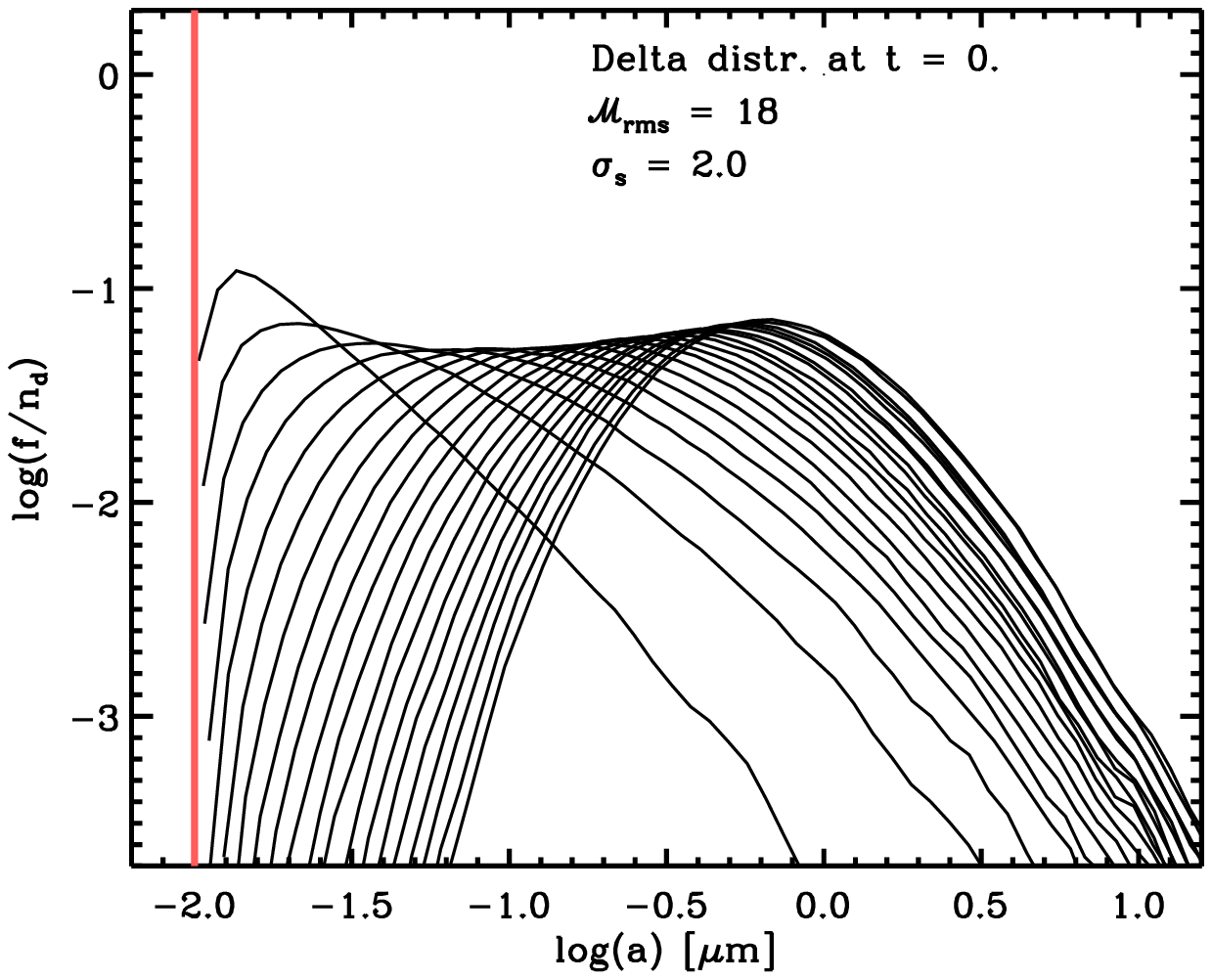}
  \includegraphics{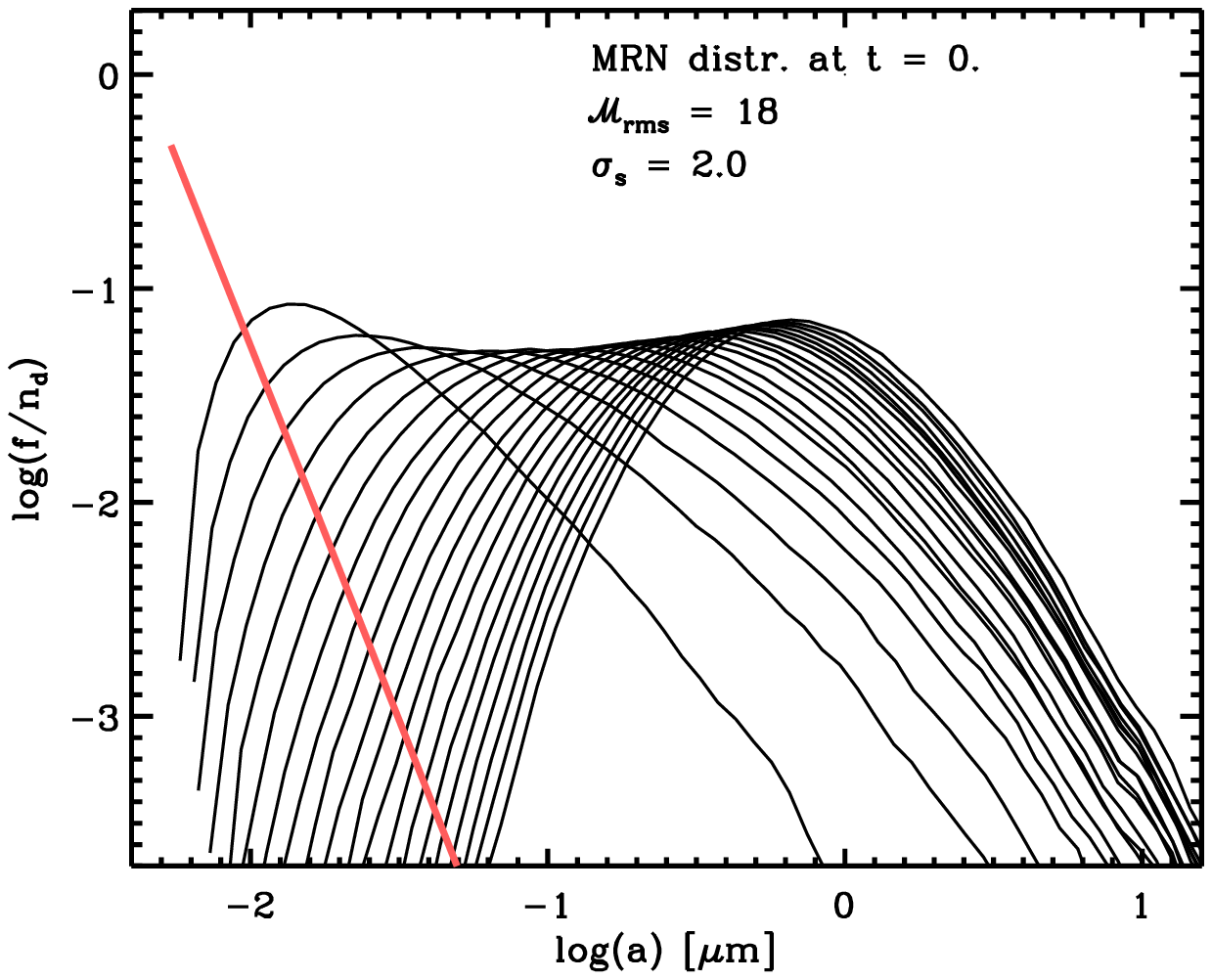}
   }
  \caption{\label{f_evol} Resultant evolution of the GSD from simulations with a mono-dispersed initial GSD (left panels) and MRN-like power-law GSD (right panels) and different values of $\sigma_s$. To show the evolution more clearly, depletion of the growth species is not taken into account in the simulations displayed here. The effect off including depletion is shown in Fig. \ref{f_evol_depl}. Due to the noisiness of the simulated GSDs at low number densities, the ordinates have been cut at $-2.4$. Hence, the full grain-size range of the initial MRN-like GSDs (right panels) is not displayed.}
  \end{figure*} 
  
  \begin{figure*}
  \resizebox{\hsize}{!}{
   \includegraphics{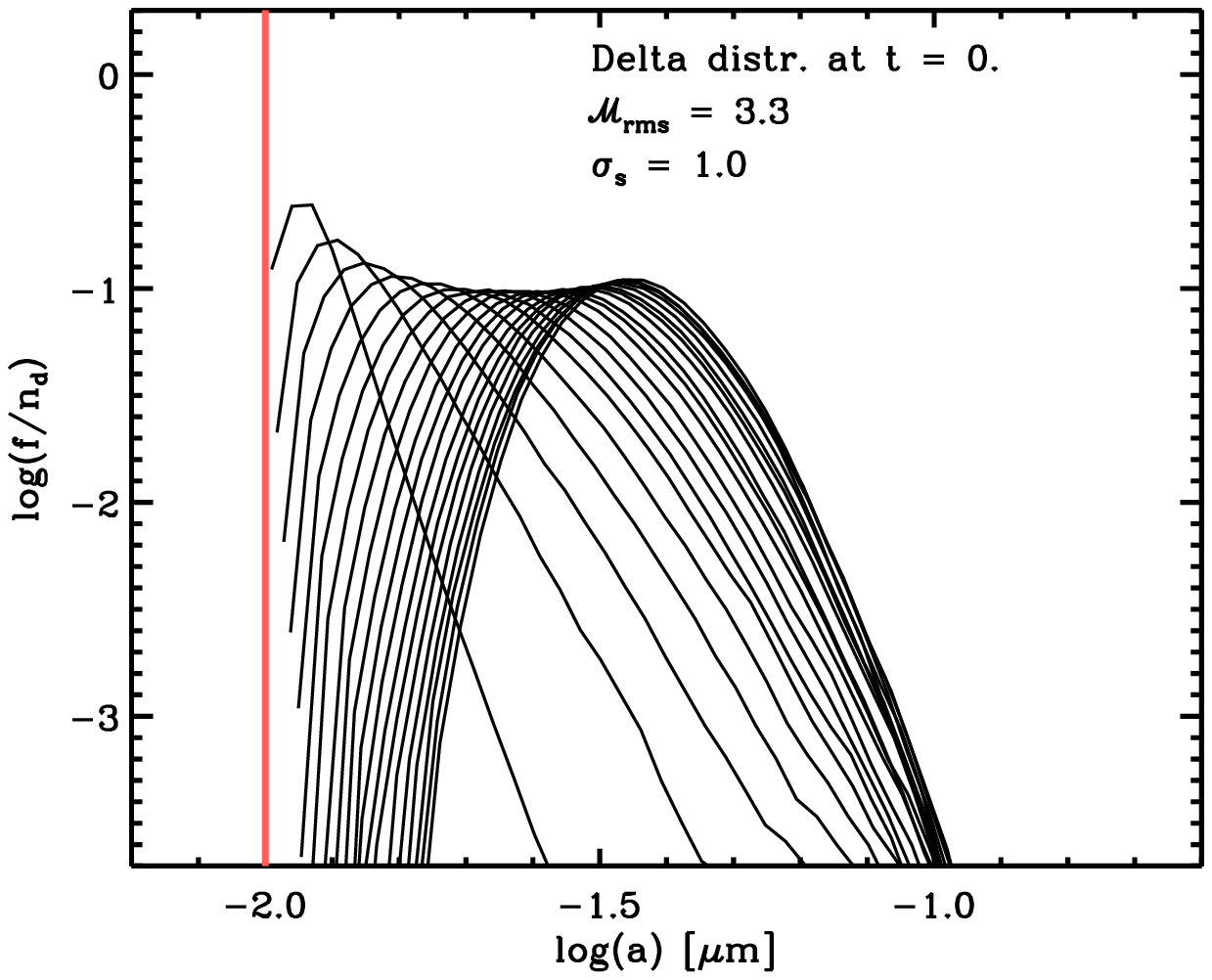}
   \includegraphics{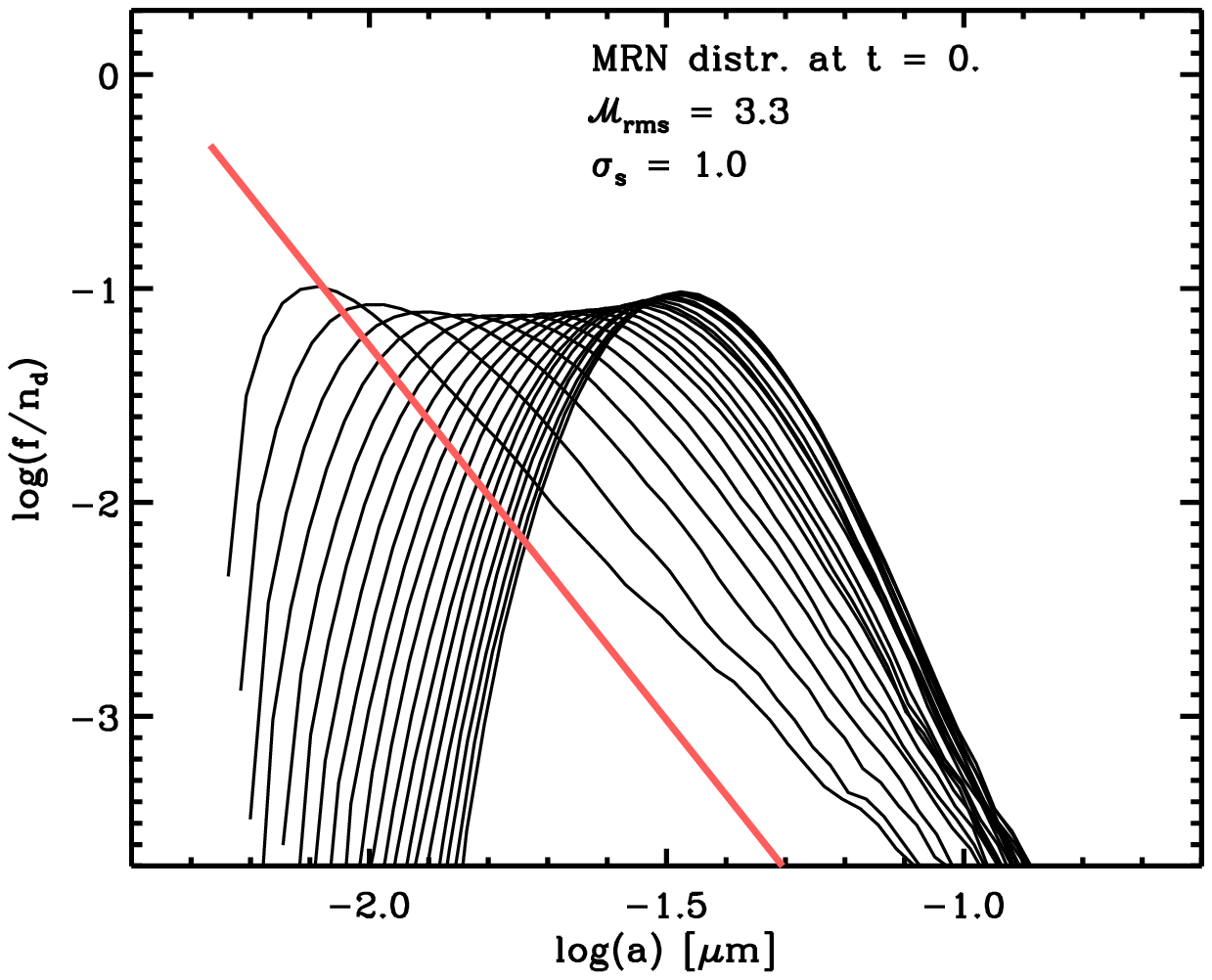}}
  \caption{\label{f_evol_depl} {Examples of GSD evolution including the effect of depletion, assuming an initial depletion of 10\%. The evolution towards a lognormal GSD is slower and may eventually stop before the GSD has obtained a truly lognormal shape. The power-law tail seen at intermediate stages may therefore remain.}}
  \end{figure*}

    \begin{figure*}
  \resizebox{\hsize}{!}{
   \includegraphics{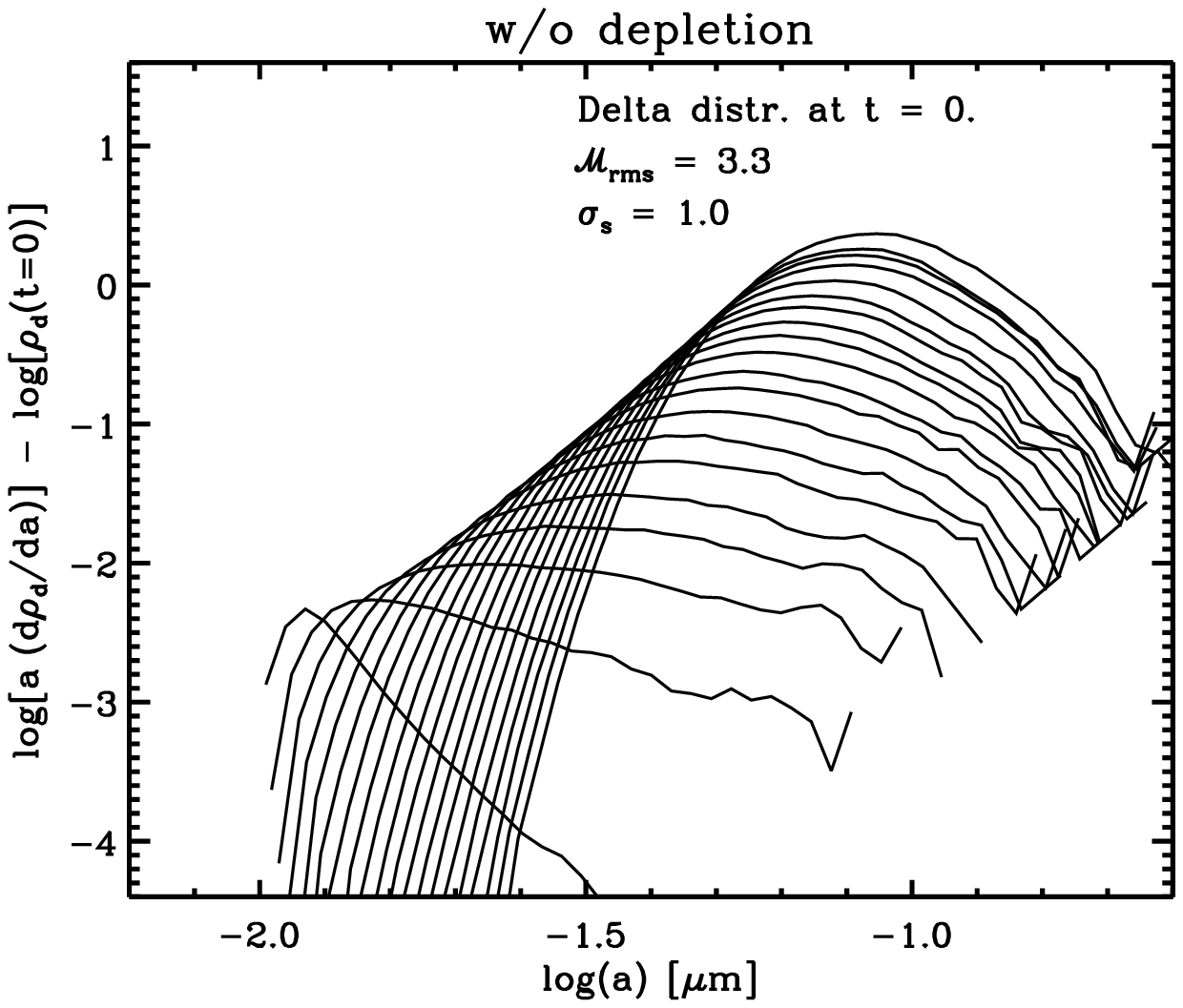}
   \includegraphics{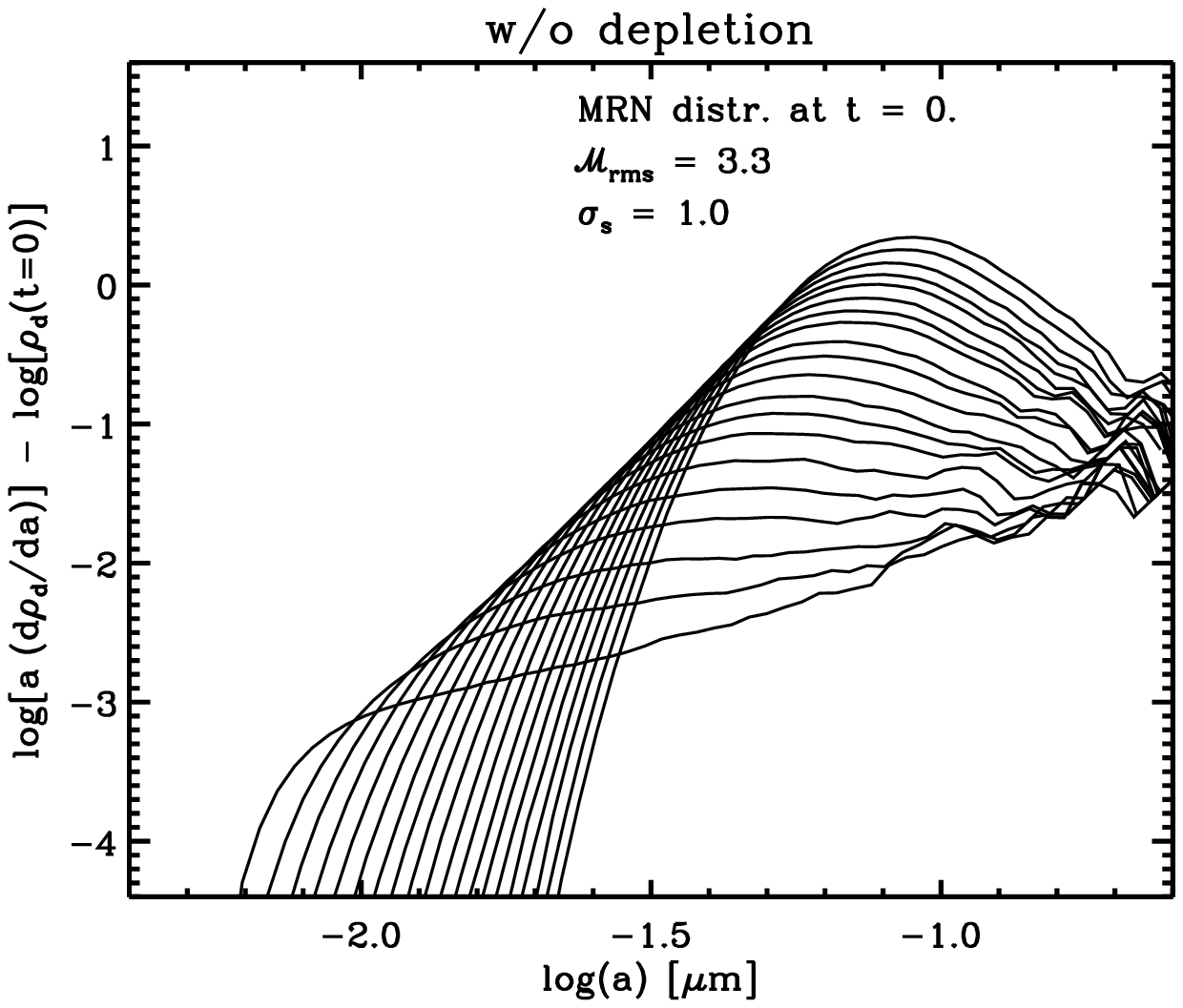}}
    \resizebox{\hsize}{!}{
   \includegraphics{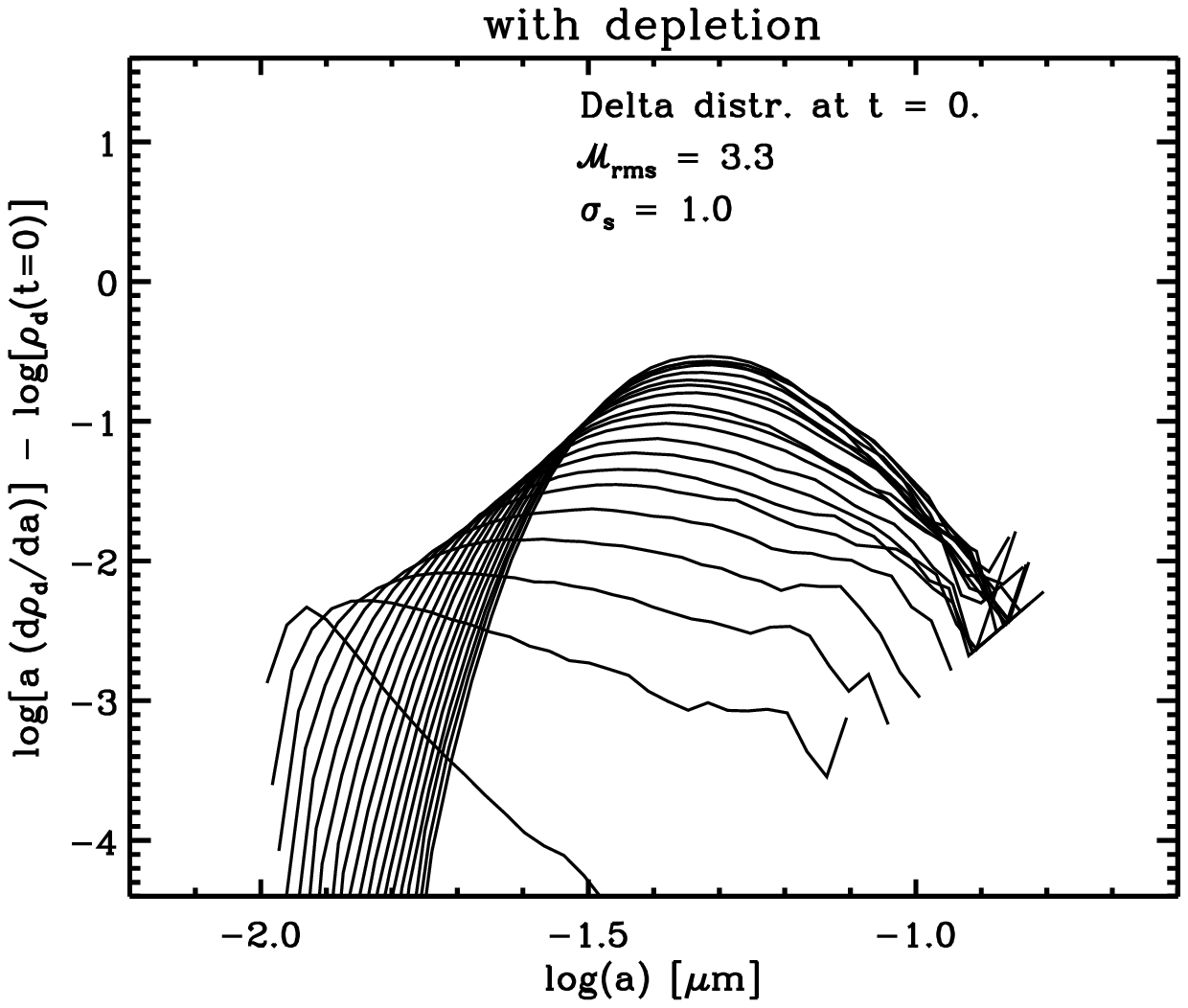}
   \includegraphics{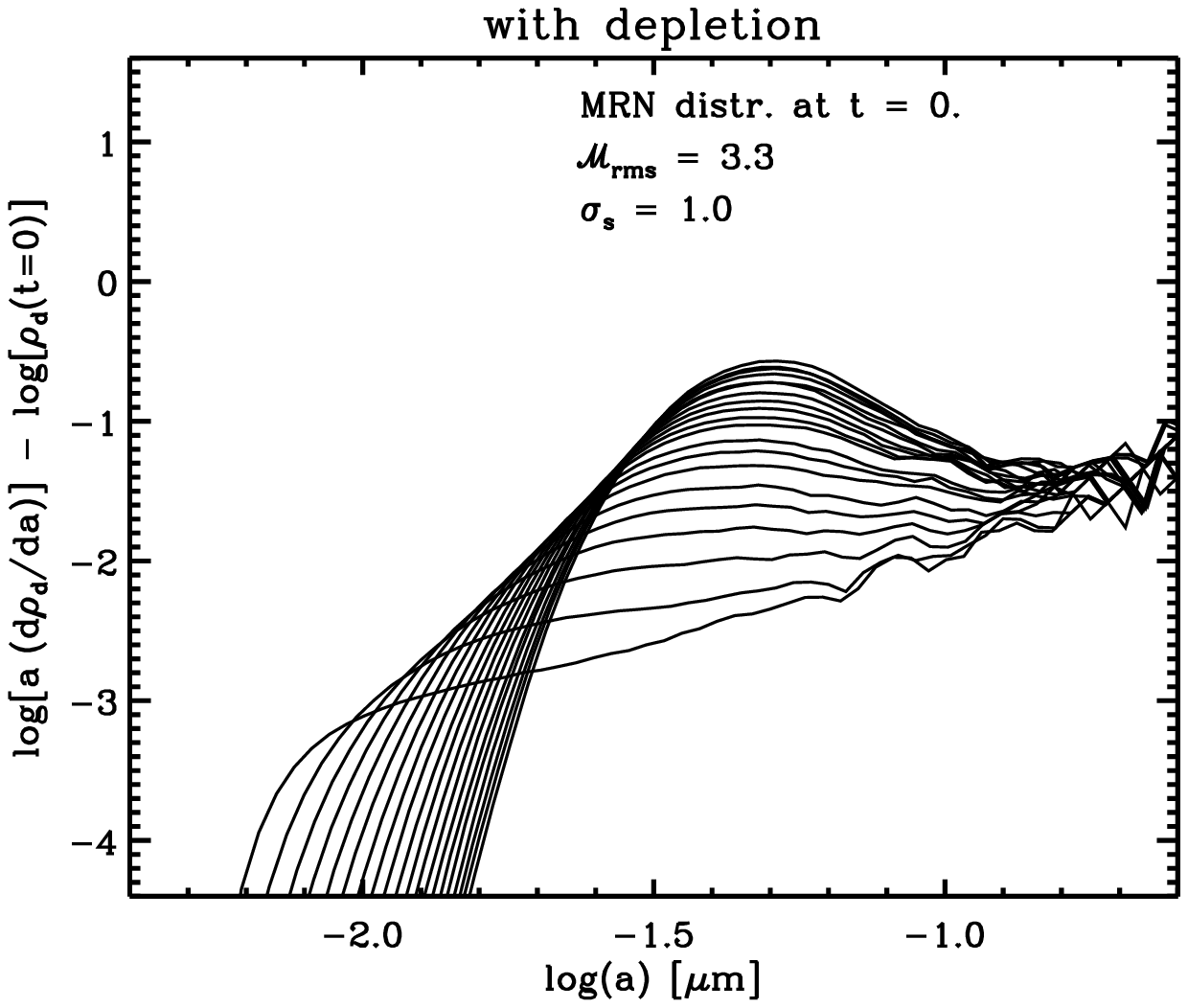}}
  \caption{\label{rhod_evol} {Evolution of the grain-mass distribution $d\rho_{\rm d}/da$ as a function of grain radius $a$. Upper panels show the case without depletion, while the lower panels show simulations with depletion. Left panels show evolution from a mono-dispersed initial GSD, while the right panels show evolution from an MRN-like initial GSD. }}
  \end{figure*}
  
  \begin{figure}
  \resizebox{\hsize}{!}{
   \includegraphics{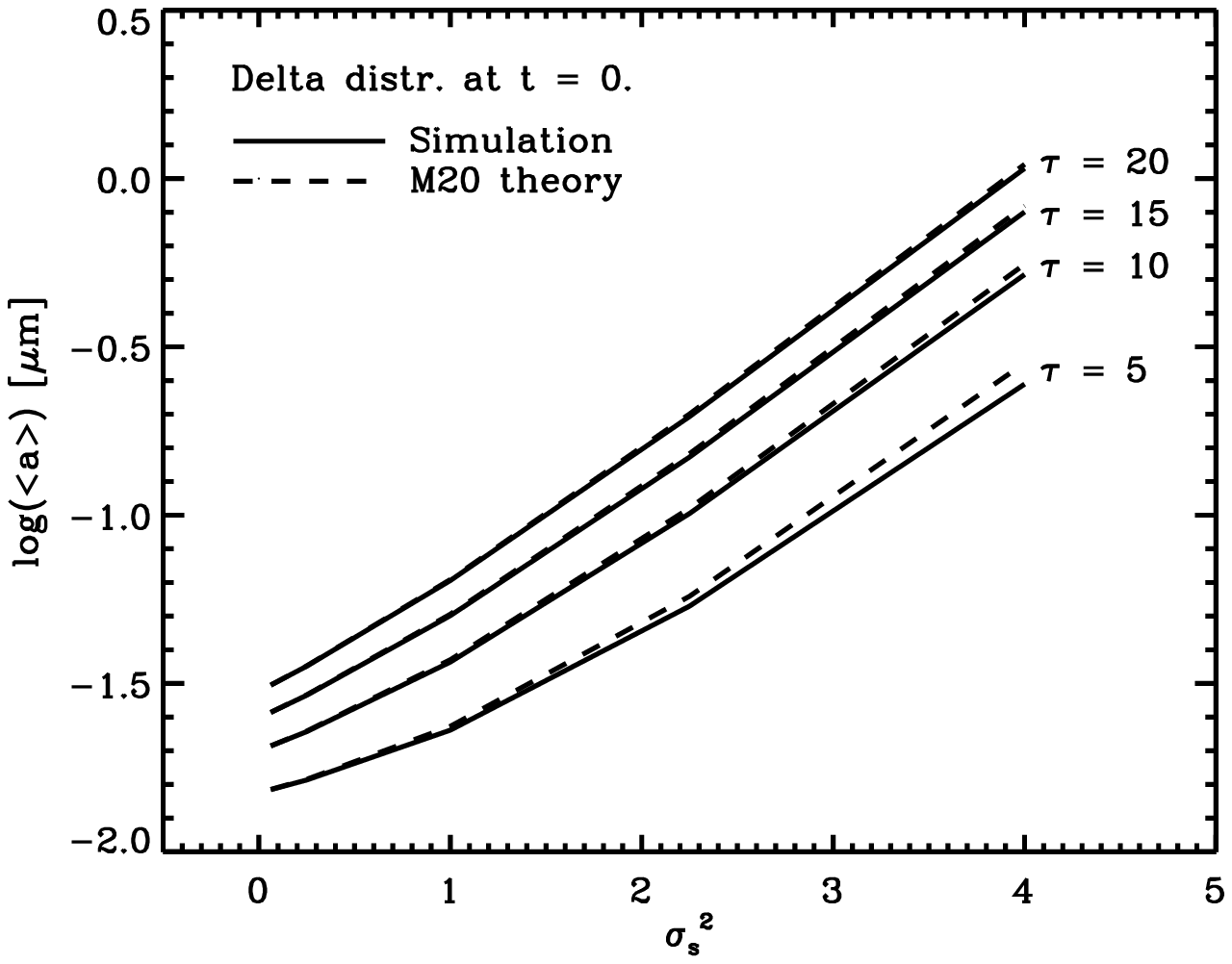}}
  \caption{\label{asigma} {Correlation between $\langle a\rangle$ and $\sigma _s$. The simulations agree well with the analytic theory of M20, in particular at later stages.}}
  \end{figure}

Fig. \ref{f_evol} shows how the GSD evolves in the different cases considered. The general trend appears to be that a power-law tail develops at early times regardless of the initial condition and after some transitional phase of temporary flattening, the evolution progresses towards a lognormal shape and a log-translational phase (see Section \ref{evolution} and Fig. \ref{fsgsd_logtrans}). It should be noted that larger values of $\sigma_s$ (higher average Mach number $\mathcal{M}_{\rm rms}$) seem to create a deviation from a lognormal tail for large $a$ in the GSDs even for the case of a mono-dispersed (delta-distributed) initial GSD. This phenomenon could in part be due to low-number statistics in the tails of the GSD, but a longer run with $N_{\rm e} = 10^6$ and $t_{\rm max} = 40$ (Run 13, see Table \ref{parameters}) still shows a power-law tail, which suggests that it could be either a real effect or an effect of the fact that the numerical method is a first-order scheme.

In general, however, there is a consistent irrefutable trend: strong turbulence (high $\sigma_s$ values) clearly leads to a wider GSD and generally larger grains. The average grain radius $\langle a\rangle$ at $t = 20$ is approximately 40 times larger for $\sigma _s = 2$ compared to $\sigma_s = 0.25$. To first order  $\log(\langle a\rangle \propto \sigma_s^2$ at any given time (see Fig. \ref{asigma}), which can be understood in terms of theoretical predictions in M20, in particular equation (17) in that paper. Note that the simulations deviate from the M20 theory at early times, which may be explained by the fact that the number of fluid elements is finite and not large enough to ensure that the simulated gas-density variation strictly follows a lognormal distribution at any given time (which is the case for the analytical theory of course).

To show the evolution towards a lognormal GSD more clearly, all simulations displayed in Fig. \ref{f_evol} are made without any correction for depletion. In reality, depletion is negligible only at early times or if the initial dust-to-gas ratio is very low. The effect of including depletion (see equation \ref{depletion}) of growth species (Run 11 and 12, see Table \ref{parameters}) is shown in Fig. \ref{f_evol_depl}, where one can see that, assuming an initial depletion of 10\%, the evolution towards a lognormal GSD is slower and may eventually stop before the GSD has obtained a truly lognormal shape. That is, the GSDs with power-law tails seen at intermediate stages can be the end states of the evolution and the log-translational phase may not occur. It is also noteworthy, that if all grain growth is depletion limited, i.e., if the key growth species is fully depleted within the lifetime of an MC, then $\langle a\rangle$ will eventually be the same regardless of what $\sigma_{\rm s}$ is, provided that all other parameters are unchanged. A central point in the M20 theory is that a state of total depletion will likely not occur within the lifetime of an MC unless the growth is accelerated by turbulence.

As dust extinction as well as emission is strongly correlated with dust mass, it is of interest to consider the dust-mass density $\rho_{\rm d}$ and how it evolves. More precisely, the quantity $a\,d\rho_{\rm d}/da$ is displaying how the total dust mass evolves and how the dust mass is distributed over the various grain sizes $a$. Fig. \ref{rhod_evol} shows this grain-mass distribution (GMD) with and without depletion for both a mono-dispersed and a MRN-like initial GSD. From Fig. \ref{rhod_evol} it is evident that the total amount of dust that can be formed is limited by depletion, which is a rather trivial result, while the overall shape of the GMD is very similar, which is a more interesting result. The GMD is initially peaking at small $a$ in case of a mono-dispersed GSD, and at large $a$ if the GSD is initially MRN-like, but in both cases the GMD evolves towards having a peak at essentially the same $a$. Obviously, depletion limited growth leads to a peak at somewhat smaller $a$; roughly a factor of two smaller than without depletion, in the present case.

\subsection{Effective grain-size dependence of $\xi$?}
As indicated above, in section \ref{evolution}, grains residing in high-density regions will grow fast and become large and as they are likely to remain in a high-density region for a while and, similarly, grains located in low-density regions will be small. Thus, the growth velocity $\xi$ will effectively depend on the grain radius $a$. More precisely, $\xi$ would increase with $a$. For simplicity, one may assume that ${\xi}(a,t) = \xi_0(t)\,(a/a_0)^{1+\beta}$, where $\beta \ne 0$. Then, by the generalised formal solution in Appendix \ref{genformalsol},
\begin{equation}
f(a,t) = {C\over\sigma_\eta\,a} \left({a\over a_0}\right)^{-n} \exp\left\{-{1\over 2\,\sigma_\eta^2}\left[{a_0\over n}\left({a\over a_0} \right)^{-n} - A(t) \right]^2 \right\},
\end{equation}
where $C = n_{\rm d}/(\sqrt{2\,\pi}\,a_0)$, $\sigma_\eta$ is the standard deviation of $\eta = a_0/n\,(a/a_0)^{-\beta} - A(t)$ and
\begin{equation}
A(t) = \int_0^t \xi_0(t^{\prime})\,dt^{\prime}.
\end{equation}
For the special case $\beta = 0$ the solution is a lognormal distribution instead. For $\beta>0$ the solution will have a power-law tail and look very similar to GSDs seen at early times in Fig. \ref{f_evol}. Thus, one may say that, effectively, the evolution of the GSD in turbulence corresponds to having a grain-size dependent $\xi$, where the dependence on the grain radius $a$ is very steep at early times and evolve towards a linear relation and a lognormal GSD.

The argument above is indeed very sketchy and phenomenological, but serves to prove that the usual assumption, that growth of spherical grains can be described by a $\xi$ which is independent of $a$, can be called into question if dynamics is not included in the model. In ``zero-dimensional'' models, without turbulent gas dynamics and resultant density variation, grain-growth by accretion of molecules will only lead to translational evolution of the GSD (see section \ref{evolution}). But such models can be modified parametrically to include the effects of turbulence by adding the $\mathcal{M}_{\rm rms}$ correction to the overall growth velocity suggested in M20 in combination with some dependence on $a$, as described above. Developing such a parametric modification in detail goes beyond the scope of the present paper, though. 

\subsection{Scaling with Mach number?}
An obvious question to ask in connection to parametric models is whether there exists some kind of scaling with the Mach number $\mathcal{M}_{\rm rms}$. Based on Fig. \ref{f_evol} it is tempting to suggest a relation between the standard deviation of the GSD $\sigma_a$ and $\mathcal{M}_{\rm rms}$ similar to the well-established relation between $\sigma_s$ and $\mathcal{M}_{\rm rms}$ (see equation \ref{sigmamach}). However, closer scrutiny shows that $\sigma_a$ and $\sigma_s$ may not have a simple functional relation. First, the initial GSD must play a role for the resultant $\sigma_a$, which can be seen in the upper panels of Fig. \ref{f_evol}, showing the two simulations with $\sigma_s = 0.25$. Second, since the evolution of the GSD is depletion limited, the initial level of depletion is also important as it determines how much grains will be able to grow, which in turn affects $\sigma_a$. Third, as mentioned above in Section \ref{numsim}, grains may grow large enough to decouple from the flow, which can affect the over all rate of growth. In this case the scaling of the problem, e.g., mean gas density $\langle\rho\rangle$ and $\sigma_s$, will be crucial. 

Given the dependencies on the initial conditions listed above, it is actually unlikely there exists a universal relation between the width/variance of the GSD and  $\mathcal{M}_{\rm rms}$. But the qualitative result, that the GSD becomes broader due to turbulence, is quite clear and the effect must be larger at high $\mathcal{M}_{\rm rms}$, regardless of initial conditions.

\subsection{Other forms of dust processing in turbulence}
The present study has focused on grain growth by accretion of molecules, but there are other forms of dust processing that may become important in turbulent MCs. In particular, growth by coagulation/aggregation is important in regions of very high density; the grain-grain interaction rate in compressible turbulence increases mainly due to over-density effects rather than turbulent velocities \citep[see][]{Li20}. Grain-grain interactions at sufficiently high energies may also lead to shattering \citep[see, e.g.,][]{Slavin04,Hirashita09}. But if coagulation dominates over shattering, the evolution of the GSD is in fact not so different from the evolution seen here due to accretion of molecules, except that small grains will always remain, creating a stretched-out GSD with a lognormal-type slope at the large-grain end \citep[see][Fig. 2]{Li20}.  The reason for this similarity between the large-grain tails is likely that while the grain growth is driven by over-densities of molecules in one case, it is driven by over-densities of small grains accreting onto large grains in the other. Small grains trace the gas quite well and therefore they follow the gas PDF just like molecular growth species.

\section{Summary and conclusions}
In a previous study (M20) it was shown that turbulence can significantly accelerate dust growth by accretion of molecules onto grains, where the growth rate scales with the square of the Mach number. Here, it is has been shown, by simulating isothermal turbulence as an Ornstein-Uhlenbeck process, how turbulence must have a significant impact also on the resultant GSD, as seen in recent hydrodynamic simulations (Li \& Mattsson 2020, submitted). In particular, the variance (``width'') of the GSD increases with the mean Mach number $\mathcal{M}_{\rm rms}$, although a generic scaling relation may not exist. 

The turbulence-induced broadening of the GSD implies that a fraction of very large grains, with radii orders of magnitudes larger than the initial mean radius, can form without any need to assume extreme conditions. For $\mathcal{M}_{\rm rms}\sim 10$ a significant fraction of micron-sized grains could form by growth by accretion only.

The shape of the GSD at later stages of evolution appears to be a reflection of of the gas-density PDF. That is, a lognormal distribution of gas densities tend to eventually produce a lognormal GSD. This corroborate the use of a lognormal GSD for large grains in ISM dust models \citep{Jones13}. Of course, the initial GSD plays a role, but the ``memory'' of the initial shape of the GSD gradually fades in all simulations presented here, while the total number density of grains is conserved. In case of highly compressible turbulence (high $\mathcal{M}_{\rm rms}$), the simulations seem to predict slightly skewed GSDs with a large-grain excess compared to simulations corresponding to lower $\mathcal{M}_{\rm rms}$, but this may be a statistical artefact.

Modelling turbulence as a stochastic process has obvious limitations, which is why the results presented above need to be confirmed with numerical simulations where one is actually solving the equations of fluid dynamics. Furthermore, the theory by \citet{Baines65} predicts that ``drift'', i.e., dynamical decoupling of dust and gas, may have an important effect on the growth rate, which is most easily explored with detailed numerical simulations. Such efforts are currently under way (Li \& Mattsson 2020, submitted).

\section*{Acknowledgments}
The author wishes to thank the anonymous reviewer, whose comments, suggestions and criticism was much appreciated.
This work is supported by the Swedish Research Council (Vetenskapsrådet), grant no. 2015-04505. 
%Nordita is financed by the Nordic Council of Ministers and the two host universities KTH Royal Institute of Technology and Stockholm University.

\section*{Data availability}
The code and corresponding output data underlying this article will be shared on reasonable request to the corresponding author.

\bibliographystyle{mnras}
\bibliography{refs_dust}

\appendix
\section{Generalisation of equation (5)}
\label{genformalsol}
Assuming that the growth velocity $\xi(a,t)$ can be separated into a time-dependent component $\xi_0(t)$ with the same unit as $\xi$ (length/time) and a non-dimensional component $B(a)$ describing the grain-size dependence, one may write the equation governing equation of the GSD as
\begin{equation}
{\partial \varphi\over \partial t}+ \xi_0(t)\,B(a)\,{\partial \varphi\over \partial a}= 0, \quad \varphi(a,t) = B(a)\,f(a,t).
\end{equation}
The formal solution to this equation can easily be derived by introducing the transformation
\begin{equation}
\eta = \int_0^a {da^{\prime}\over B(a^{\prime})} -\int_0^t \xi_0(t^{\prime})\,dt^{\prime},
\end{equation} 
which simply yields that $\varphi(\eta) = \mathcal{F}(\eta)$, where $\mathcal{F}$ is an arbitrary function that can be defined by initial or end-state conditions. Thus, if $\mathcal{F}(\eta) \propto e^{-{1\over 2} \eta^2}$, i.e., the transformation variable $\eta$ follows a zero-mean Gaussian distribution if $\xi$ has no dependence on $a$ ($B = 1$), leads to
\begin{equation}
\varphi(a,t) = {C \over\sigma_\eta}\exp\left[-{1\over 2\,\sigma_\eta^2}\left( \int_0^a {da^{\prime}\over B(a^{\prime})} -\int_0^t \xi_0(t^{\prime})\,dt^{\prime}\right)^2 \right],
\end{equation} 
where $n_{\rm d}$ is the number density of grains, $\sigma_\eta$ is the standard deviation of $\eta$, $C = n_{\rm d}/(\sqrt{2\,\pi}\,a_0)$ and $a_0$ is the initial mean grain radius. The special case $B(a) = a/a_0$ (equation \ref{eoc_log}) would then correspond to a lognormal GSD.  

\label{lastpage}
\end{document}